\documentclass[conference]{IEEEtran}

\usepackage{graphicx}
\usepackage{amsmath,amssymb,amsthm}

\newtheorem{theorem}{Theorem}

\begin{document}

\title{Distributed Cloud Association in Downlink Multicloud Radio Access Networks}

\author{
Hayssam Dahrouj, Tareq Y. Al-Naffouri and Mohamed-Slim Alouini\\
Computer, Electrical and Mathematics Sciences and Engineering Division\\ King Abdullah University of Science and Technology (KAUST), Thuwal, Makkah Province, Saudi Arabia \\
E-mails: \{hayssam.dahrouj,tareq.alnaffouri,slim.alouini\}@kaust.edu.sa}


\maketitle

\begin{abstract}

This paper considers a multicloud radio access network (M-CRAN), wherein each cloud serves a cluster of base-stations (BS's) which are connected to the clouds through high capacity digital links. The network comprises several remote users, where each user can be connected to one (and only one) cloud. This paper studies the user-to-cloud-assignment problem by maximizing a network-wide utility subject to practical cloud connectivity constraints. The paper solves the problem by using an auction-based iterative algorithm, which can be implemented in a distributed fashion through a reasonable exchange of information between the clouds. The paper further proposes a centralized heuristic algorithm, with low computational complexity. Simulations results show that the proposed algorithms provide appreciable performance improvements as compared to the conventional cloud-less assignment solutions.
\end{abstract}


\IEEEpeerreviewmaketitle

\thispagestyle{empty}

\section{Introduction}

Cloud radio access networks (CRAN) are expected to be the core new network architecture in next generation mobile radio systems \cite{Andrews_Buzzi_Choi_Hanly_Lozano_Soong_Zhang}. To support the ever increasing demand for high-speed data, base-stations are increasingly deployed in smaller cell sizes with a progressive move towards full spectrum reuse. By connecting the numerous base-stations via high-speed links to centralized cloud computing processors, CRAN provides an efficient cellular architecture that enables large-scale interference management through coordinated and joint signal processing. While the majority of recent works focus on a single-cloud scenario and neglects intercloud interference, this letter considers the more favorable practical multicloud scenario and addresses the user-to-cloud assignment problem.

The model considered in this paper is a practical realization of a CRAN system over a dense multicell network. It consists of a radio access network comprising several clouds, as opposed to the single-cloud scenario assumed in the recent CRAN literature, e.g., see \cite{Andrews_Buzzi_Choi_Hanly_Lozano_Soong_Zhang} and references therein. A multicloud model is recently considered in \cite{Park_Simeone_Sahin_Shamai_WCL_letter}; however, the problem addressed in \cite{Park_Simeone_Sahin_Shamai_WCL_letter} is based on a pre-known association of clouds and users. The user-to-cloud assignment problem studied in this paper is also related to the base-station association problem which is well studied in the literature of wireless networks. However, the majority of the previous works either focus on centralized solutions to the problem \cite{Han_Farrokhi_Ji_Liu}, or derive distributed solutions for specific utilities, e.g. log-rate maximization \cite{Shen_Yu_JSAC}. Most importantly, the methods in the multiple-input-multiple-output (MIMO) base-station scenario are unsatisfactory in the distributed antenna infrastructure supported in CRANs, as the base-stations connected to one CRAN are not co-located, and so the channel cannot be simply averaged over different paths as in \cite{Shen_Yu_JSAC}.

This paper formulates an optimization problem that maximizes a \textit{generic network-wide utility function} subject to network connectivity constraints where each user cannot be connected to more than one cloud at a time, and each cloud operates according to a resource budget constraint, e.g., the number of users each cloud serves cannot exceed the number of base-sations' antennas connected to the cloud, so as to preserve high system multiplexing gain. The problem is formulated as a generalized assignment problem (GAP), which is an NP-hard problem. The majority of the available solutions in the literature of GAP, both in computer science and operational research, are centralized in nature, e.g. \cite{Shmoys_Tardos}. The main contribution of this paper is that it solves the multicloud association problem using an iterative auction approach, first proposed in \cite{Luo_Chakraborty_Sycara}, utilizing a knapsack-subroutine \cite{knapsack_book}. The proposed method can be implemented in a distributed fashion across the multicloud network, and only requires a reasonable amount of information exchange between the clouds. The paper further proposes a centralized heuristic algorithm with low computational complexity. Simulations results show that the proposed algorithms provide appreciable performance improvements as compared to the conventional cloud-less assignment solutions.

\section{System Model and Problem Formulation}

\subsection{System Model}
Consider the downlink of a multicloud radio access network, composed of $C$ clouds each
serving $B$ basestations, over a network comprising $U$ users. The base-stations are assumed to be connected to the clouds via high-capacity digital links. We further assume that base-stations and users are equipped with single antennas.

Let $\mathcal{C}=\{1,\cdots,C\}$ denote the set of clouds, and $\mathcal{U}=\{1,\cdots,U\}$ be the set of users. Each user $u\in \mathcal{U}$ can be assigned to one and only one cloud  $c \in \mathcal{C}$. Furthermore, every cloud $c \in \mathcal{C}$ has its own resource budget constraint, e.g., the constraint on the number of users that it can be connected to.

Let $h_{cbu}\in {\mathbb C}$ be the channel from the $b$th BS of the $c$th cloud to the $u$th user, and let ${\textbf{h}}_{cu}\in {\mathbb C}^{B\times 1}$ be the channel vector from the $c$th cloud to the $u$th user, i.e., ${\textbf{h}}_{cu}= [{{h}}_{c1u},\cdots, {{h}}_{cBu}]^T$. Define $\textbf{w}_{cu} \in {\mathbb C}^{B\times 1}$ be the transmit beamformer over cloud $c$'s BS's for user $u$, which is fixed throughout this paper.

\subsection{Problem Formulation}
Let $r_{cu}$ be the generic reward of associating user $u$ to cloud $c$, and $A_{cu}$ be the binary association variable which is equal to 1 if user $u$ is associated to cloud $c$, and zero otherwise. We focus on solving the generalized cloud-association problem, where each user can be connected to one cloud at most, and where every cloud has it own resource connectivity constraint. The paper considers the following network-wide optimization problem:

\begin{eqnarray}
\label{generalized_optimization_problem}
& \max & \sum_{c,u}r_{cu}A_{cu} \\
& {\rm s.t.\ } & \sum_{c\in \mathcal{C}}A_{cu}\leq 1,\quad \forall u\in \mathcal{U}\nonumber\\
& & \sum_{u\in \mathcal{U}}\alpha_{cu}A_{cu}\leq K_c,\quad \forall c\in \mathcal{C}\nonumber\\
& & A_{cu}\in \{0,1\}, \forall (c,u)\in \mathcal{C}\times \mathcal{U}, \nonumber
\end{eqnarray}
where the optimization is over the binary variable $A_{cu}$, and where the constraint $\sum_{u\in \mathcal{U}}\alpha_{cu}A_{cu}\leq K_c$ denotes the resource connectivity constraint of cloud $c$. For example, $\alpha_{cu}=1$ and $K_c=B$ physically mean that cloud $c$ spatially multiplexes $B$ users at most, since the cluster of base-stations served by cloud $c$ behaves as one distributed antenna system of $B$ antennas.

%

This paper focuses on solving problem (\ref{generalized_optimization_problem}) by assuming that, at this stage, once a user is associated with a cloud, it is served by all base-stations of that cloud at a fixed power transmission. In a nutshell, all beamforming vectors $\textbf{w}_{cu}$ are fixed throughout this paper. More precisely, each beamforming vector $\textbf{w}_{cu}$ is set to the ones-vector scaled by some fixed power value $P_c$, $\forall$ $c\in \mathcal{C}$. The insight for such assumption is that, in the downlink, calculating the benefit of associating user $u$ to cloud $c$ becomes independent of the association of other users across the network. Finding the index and power value of base-stations serving each user within each cloud is an extra stage which eventually corrects for the appropriate beamforming vectors, e.g., \cite{Shen_Yu_JSAC}. But the second stage falls outside the scope of this paper.

%

\section{User-to-Cloud Association}
This section proposes a distributed algorithm to solve the cloud-association problem (\ref{generalized_optimization_problem}). It is based on the iterative auction-based approach presented in \cite{Luo_Chakraborty_Sycara}. The algorithm called distributed cloud-association algorithm (DCAA) can be implemented in a distributed fashion across the network. The paper further presents a centralized heuristic algorithm with low computational complexity.

\subsection{Distributed Cloud-Association Algorithm (DCAA)}
The main idea in DCAA is that each cloud $c$ bids for users which maximize cloud $c$'s net benefit, taking into consideration the penalty-tag $\lambda_u$, which can be seen as the price of being associated with a certain user $u$. The net benefit of cloud $c$ if it is assigned to user $u$ becomes $\pi_{cu}=r_{cu}-\lambda_u$. Each cloud $c$ strives to be assigned to users that maximize its overall net benefit: $ \sum_{u}\pi_{cu} A_{cu}$. The algorithm iteratively proceeds in updating the assignment of each cloud, in view of the other clouds' assignment.

\subsubsection{DCAA Description}
At each iteration $t$, let $\lambda_u(t-1)$ be the starting price that has to be paid for a cloud to be assigned to user $u$. The net benefit from assigning cloud $c$ to user $u$ becomes $\pi_{cu}(t-1)=r_{cu}-\lambda_u(t-1)$. Cloud $c$, then, bids for users which maximize its overall net benefit. In other terms, at iteration $t$, cloud $c$ solves the following optimization problem:
\begin{eqnarray}
\label{knapsack_optimization_problem}
& \max & \sum_{u}\pi_{cu}(t-1)A_{cu} \\
& {\rm s.t.\ }& \sum_{u\in \mathcal{U}}\alpha_{cu}A_{cu}\leq K_c \nonumber\\
& & A_{cu}\in \{0,1\}, \forall u\in \mathcal{U}, \nonumber
\end{eqnarray}
where the maximization is over the binary variable $A_{cu}$. Problem (\ref{knapsack_optimization_problem}) is a knapsack problem, which is an NP-hard problem. There exists, however, a fully polynomial time approximation scheme which finds the optimal approximate solution of (\ref{knapsack_optimization_problem}) to any specified degree and outputs the set of users $\mathcal{U}_{c}(t)$; see \cite{knapsack_book} and references therein.

After solving (\ref{knapsack_optimization_problem}), cloud $c$ bids for the set of users in $\mathcal{U}_{c}(t)$ and updates their prices as $\lambda_u(t)=r_{cu}$. Since $\mathcal{U}_{c}(t)$ solves the maximization problem (\ref{knapsack_optimization_problem}), we have: $\forall u \in \mathcal{U}_{c}(t)$, $r_{cu}-\lambda_u(t-1)> 0$. Setting $\lambda_u(t)=r_{cu}$, therefore, guarantees the increase of the price of user $u$, i.e. $\lambda_u(t)>\lambda_u(t-1)$. Such increase in the price of user $u$ makes user $u$ less favorable to clouds $c', \forall c'\neq c $ in the next iteration $(t+1)$. Note that after running the algorithm for all clouds, if user $u$ remains among the set of users associated with cloud $c$, the price of user $u$ is reset to zero before solving problem (\ref{knapsack_optimization_problem}) for cloud $c$, so that the net benefit of associating user $u$ to cloud $c$ is at its maximum $r_{cu}$, after user $u$ shows a mutual interest in cloud $c$.

At iteration $t$, without loss of generality, consider cloud $c=\mod(t-1,C)+1$, where $\mod(.,.)$ represents the modulo operator which simply allows to iterate over all clouds in a sequential manner as the iterations index increases. Let $\beta_{cu}(t)$ denote the bids of cloud $c$ to users $u\in \mathcal{U}_{c}(t)$. The algorithm described above, called DCAA, can be summarized as follows:



\begin{enumerate}
\item Set the iteration index $t=1$, and the initial set of users's prices $\lambda_u(0)=0, \forall u\in \mathcal{U}$.
\item At each iteration $t$, consider cloud $c=\mod(t-1,C)+1$.
\item If $t\leq C$, go to step 5.
\item If $t>C$, $\forall u\in \mathcal{U}_{c}(t-C)$, reset the prices of users that are still associated with cloud $c$, i.e., if there exist some users $u\in \mathcal{U}_{c}(t-C)$ such that $\lambda_u(t-1)=\beta_{cu}(t-C)$, then set their prices $\lambda_u(t-1)$ to zero.
    \item Calculate the net benefits $\pi_{cu}(t-1)=r_{cu}-\lambda_u(t-1)$, and solve the knapsack problem of cloud $c$, i.e. problem (\ref{knapsack_optimization_problem}), which determines the updated set of users associated with cloud $c$, denoted by $\mathcal{U}_{c}(t)$:
\begin{enumerate}
\item $\forall u\in \mathcal{U}_c(t)$, update the bids of cloud $c$ to users $u$ as $\beta_{cu}(t)=r_{cu}$, and the prices of users $ u\in \mathcal{U}_c(t)$ to $\lambda_u(t)=\beta_{cu}(t)$.
 \item $\forall u\notin \mathcal{U}_c(t)$, keep the prices unchanged, i.e., $\lambda_u(t)=\lambda_u(t-1)$.
\end{enumerate}
\item Set $t=t+1$; go to step 2; and stop at convergence.
\end{enumerate}


\begin{theorem}
The iterative auction-based algorithm DCAA is guaranteed to converge in a finite number of iterations with an approximation ratio $(1+\gamma)$, where $\gamma\in [1,+\infty)$ is the approximation ratio of the subroutine knapsack algorithm used in step 5 above. In other terms, the solution reached by DCAA, denoted $f^{DCAA}$, is $(1+\gamma)$ away from the global optimal solution $f^*$: $(1+\gamma)f^{DCAA}\geq f^*$.
\end{theorem}
Steps for the proof of theorem 1 are omitted in this paper as they mirror theorem 1 and theorem 2 of \cite{Luo_Chakraborty_Sycara}.

\subsubsection{Distributed Implementation}

To implement DCAA at iteration $t$, cloud $c=\mod(t-1,C)+1$ utilizes the set of prices $\lambda_u(t-1)$, the set of benefits $r_{cu}$, the set of weights $w_{cu}$, the set of users $\mathcal{U}_{c}(t-C)$ associated with cloud $c$ at iteration $(t-C)$, and the set of bids $\beta_{cu}(t-C)$ of cloud $c$ for users $u \in \mathcal{U}_{c}(t-C)$.

$r_{cu}$, $w_{cu}$, $\mathcal{U}_{c}(t-C)$, and $\beta_{cu}(t-C)$ are all available at cloud $c$. The set of prices $\lambda_u(t-1)$ set during iteration $t-1$ is the output of cloud $c'$ operation, where $c'=\mod(t-2,C)+1$. A distributed implementation of DCAA is, therefore, possible by a reasonable and simple exchange of users' prices from cloud $c'$ to cloud $c$.

\subsection{Centralized Heuristic Cloud-Association Algorithm (CHCAA)}
DCAA solves the cloud-association problem (\ref{generalized_optimization_problem}) using the knapsack routine which is NP-hard in general. This section presents an alternative low complexity, yet centralized, heuristic to solve (\ref{generalized_optimization_problem}). The method, denoted by centralized heuristic cloud-association algorithm (CHCAA), associates users to clouds based on the individual utilities $r_{cu}$. Let $\bf{R}$ be the $C\times U$ matrix whose entries are the potential individual utilities $r_{cu}$, i.e., the $(c,u)$th entry of the matrix $\bf{R}$ is ${\bf{R}}_{c,u}=r_{cu}$.

At each step, find the largest entry of the matrix $\bf{R}$, call it ${\bf{R}}_{c^{max},u^{max}}$. User $u^{max}$ then maps to cloud $c^{max}$, as long as the resource constraint of cloud $c^{max}$ is still satisfied. Once user $u^{max}$ gets associated with a certain cloud, delete the column of $\bf{R}$ containing ${\bf{R}}_{c^{max},u^{max}}$, so that user $u^{max}$ cannot be connected to other clouds in subsequent steps. Repeat the above procedure and stop when all users are associated with one cloud each, or when all clouds' resource constraints are violated with the addition of one more user. As the simulations results suggest, DCAA and CHCAA show a similar performance, and they both outperform conventional systems using the classical cloud-less assignment solutions.


\section{Simulations}

This section evaluates the performance of the proposed methods in a 7-cell CRAN network, which comprises $C=7$ clouds, $B=3$ base-stations per cloud, and several users distributed across the network.  The clouds are located at the center of each cell, and the distance between adjacent clouds is varied in the simulations. The simulations consider the sum-rate maximization problem, i.e. $r_{cu}=\log_2(1+\text{SINR}_{cu})$ where $\text{SINR}_{cu}$ is the signal-to-interference plus noise ratio of user $u$ when associated with cloud $c$. Further, For illustration, we choose $\alpha_{cu}=1$ and $K_c=B$ $\forall (c,u)$, so as to impose the constraint that each cloud can multiplex $B$ users at most.
\begin{figure}
\begin{center}
\rotatebox{0}{\scalebox{0.4}{\includegraphics{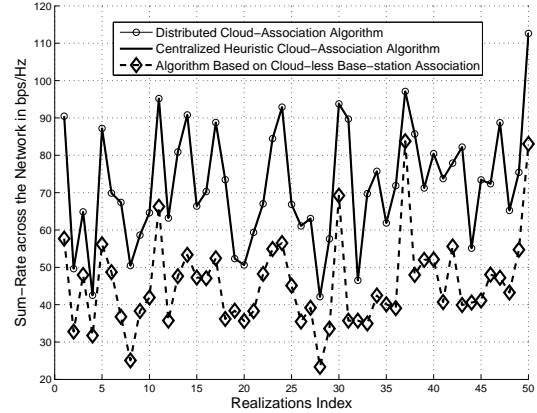}}}
\caption{Sum-rate in bps/Hz for a different number of realizations over a network comprising 7 clouds and 3 base-stations per cloud. Total number of users is 28 users, and the intercell distance is 0.5 km.} \label{DCAA_CHCAA_nrealizations}
\end{center}
\end{figure}
Fig.~\ref{DCAA_CHCAA_nrealizations} illustrates the sum-rate performance of the proposed cloud-association algorithms in bps/Hz for different channel realizations, for a network comprising 28 users where the intercell distance is set to 0.5 km. The figure shows that both the distributed cloud-association algorithm (DCAA) and the centralized heuristic cloud-association algorithm (CHCAA) have a similar performance. The difference between the two is that CHCAA has a low computational complexity as compared to DCAA which is an iterative algorithm involving a knapsack solution at each iteration. DCAA, on the other hand, can be implemented in a distributed fashion across the different clouds. Fig.~\ref{DCAA_CHCAA_nrealizations}, further, shows how both DCAA and CHCAA outperform the cloud-less base-station association solution for all realizations of the channel, which highlights the importance of using clouds for associating users in CRAN networks.

\begin{figure}
\begin{center}
\rotatebox{0}{\scalebox{0.4}{\includegraphics{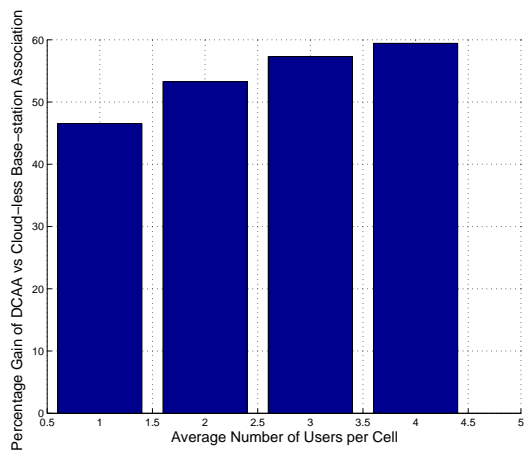}}}
\caption{Percentage gain in sum-rate of the proposed algorithm as compared to base-station association in the absence of clouds for different number of users. The network comprises 7 clouds and 3 base-stations per cloud. The intercell distance is 0.5 km.} \label{Gain_of_DCAA_vs_BS_association}
\end{center}
\end{figure}

To illustrate the gain of the cloud-association algorithms as a function of the number of users, Fig.~\ref{Gain_of_DCAA_vs_BS_association} shows the percentage gain in sum-rate for DCAA as compared to the cloud-less base-station association, for a network of 0.5 km intercell distance. As shown in the figure, when the number of users increases, the performance gain due to cloud association increases and reaches up to 60\% improvement when the average number of users per cell is 4 (i.e. total number of users is 28). Such increase in gain is due to the fact that for a larger number of users, interference becomes higher, and so the role of cloud-association as an interference mitigation technique becomes more pronounced.

%
\section{Conclusions}
Optimization in cloud-radio access networks is a topic of significant interest for emerging wireless networks. The paper utilizes an auction-based iterative algorithm to solve the cloud-association problem. The algorithm can be implemented in a distributed fashion across the multiple clouds using using a reasonable amount of information exchange between the clouds. The paper further proposes a centralized heuristic algorithm with low computational complexity.

\section*{Acknowledgements}
The authors wish to thank Lingzhi Luo from Carnegie Mellon University for his support and helpful discussions.

\bibliographystyle{IEEEtran}
\bibliography{IEEEabrv,reference}

\end{document}